\documentclass[a4paper,11pt]{article}
\pdfoutput=1 

\usepackage[utf8]{inputenc}
\usepackage{jheppub}
\usepackage{amsmath}
\usepackage{amsfonts}
\usepackage{amssymb}
\usepackage{mathtools}
\usepackage{amsthm}
\usepackage{dsfont}
\usepackage{hyperref}
\usepackage[capitalise,sort]{cleveref}
\usepackage{pgfplots}
\usepackage{tabularx}
\usepackage{environ}
\usepackage[inline]{enumitem}
\usepackage{comment}
\usepackage{tikz}
\usetikzlibrary{decorations.pathmorphing,calc,fit,matrix,cd,shapes,arrows,positioning, intersections,chains,positioning,arrows.meta,hobby,patterns,shadings,shadows}
\usepackage{blkarray}
\usepackage{shadow}
\usepackage{fancybox}
\usepackage{bm}
\usepackage{lscape}
\usepackage{verbatim}
\usepackage[normalem]{ulem}
\usepackage{mdframed}
\usepackage{orcidlink}
\usepackage{subfig}

\pgfplotsset{
	compat=1.9,
	compat/bar nodes=1.8,
}

\makeatletter
\def\@xfootnote[#1]{%
	\protected@xdef\@thefnmark{#1}%
	\@footnotemark\@footnotetext}
\makeatother

\definecolor{prhigh}{HTML}{ff0000}
\definecolor{sechigh}{HTML}{e0fbfc}
\definecolor{prcolor}{HTML}{1d3557}
\definecolor{seccolor}{HTML}{457b9d}
\definecolor{tercolor}{HTML}{98c1d9}
\definecolor{blueplot}{HTML}{58468e}

\newcommand\bea{\begin{eqnarray}}
	\newcommand\eea{\end{eqnarray}}

\theoremstyle{plain}

\theoremstyle{definition}

\newtheorem*{conjecture*}{Conjecture}

\newtheorem{remark*}{Remark}



\DeclareMathOperator{\U}{U}

\newcommand{\de}{\partial}

\newcommand{\ID}{\mathds{1}}

\newcommand{\coma}{\, , \quad}
\newcommand{\fstop}{\, .}

\newcommand{\AdS}{\text{AdS}}

\newcommand{\KK}{\text{\tiny KK}}

\newcommand{\PlD}[1]{\text{\tiny Pl,\,#1}}

\newcommand{\UoD}[1]{\text{\tiny U(1),\,#1}}
\newcommand{\ttiny}[1]{\text{\tiny #1}}

\newcommand{\AdSD}[1]{\text{\tiny AdS,\,#1}}

\newcommand{\fq}{\mathfrak{q}}

\newcommand{\fm}{\mathfrak{m}}

\renewcommand{\epsilon}{\varepsilon}

\makeatletter
\newsavebox{\measure@tikzpicture}
\NewEnviron{scaletikzpicturetowidth}[1]{%
	\def\tikz@width{#1}%
	\begin{lrbox}{\measure@tikzpicture}%
		\BODY
	\end{lrbox}%
	\pgfmathparse{#1/\wd\measure@tikzpicture}%
	\BODY
}
\makeatother

\makeatletter
\newcommand{\inlineitem}[1][]{%
	\ifnum\enit@type=\tw@
	{\descriptionlabel{#1}}
	\hspace{0pt}%
	\else
	\ifnum\enit@type=\z@
	\hspace{-15pt} \refstepcounter{\@listctr}\fi
	\quad\@itemlabel\hspace{0pt}%
	\fi}
\makeatother
\DeclareMathAlphabet{\mathdutchcal}{U}{dutchcal}{m}{n}

\tikzset{
	partial ellipse/.style args={#1:#2:#3}{
		insert path={+ (#1:#3) arc (#1:#2:#3)}
	}
}

\tikzset{cross/.style={cross out, draw=black, fill=none, minimum size=2*(#1-\pgflinewidth), inner sep=0pt, outer sep=0pt}, cross/.default={2pt}}

\tikzset{
	pics/torus/.style n args={3}{
		code = {
			\providecolor{pgffillcolor}{rgb}{1,1,1}
			\begin{scope}[
				yscale=cos(#3),
				outer torus/.style = {draw,line width/.expanded={\the\dimexpr2\pgflinewidth+#2*2},line join=round},
				inner torus/.style = {draw=pgffillcolor,line width={#2*2}}
				]
				\draw[outer torus] circle(#1);\draw[inner torus] circle(#1);
				\draw[outer torus] (180:#1) arc (180:360:#1);\draw[inner torus,line cap=round] (180:#1) arc (180:360:#1);
			\end{scope}
		}
	}
}

\tikzset{
	pics/hole/.style n args={2}{
		code = {
			\draw[fill=white] (0,0) arc(120:60:#1 and #2)  arc(-60:-120:#1 and #2);
			\draw (0,0) arc(-120:-130:#1 and #2) (#1,0) arc(-60:-50:#1 and #2);
		}
	}
}

\makeatletter
\newcommand*{\itemequation}[3][]{%
	\item
	\begingroup
	\refstepcounter{equation}%
	\ifx\\#1\\%
	\else  
	\label{#1}%
	\fi
	\sbox0{#2}%
	\sbox2{$\displaystyle#3\m@th$}%
	\sbox4{\@eqnnum}%
	\dimen@=.5\dimexpr\linewidth-\wd2\relax
	\ifcase
	\ifdim\wd0>\dimen@
	\z@
	\else
	\ifdim\wd4>\dimen@
	\z@
	\else 
	\@ne
	\fi 
	\fi
	\@latex@warning{Equation is too large}%
	\fi
	\noindent   
	\rlap{\copy0}%
	\rlap{\hbox to \linewidth{\hfill\copy2\hfill}}%
	\hbox to \linewidth{\hfill\copy4}%
	\hspace{0pt}
	\endgroup
	\ignorespaces 
}
\makeatother  

\hypersetup{
	pdftitle={Expanding the Weak Gravity Conjecture in AdS Space},    
	pdfauthor={\textcopyright\ Puxin Lin, Alessandro Mininno, Gary Shiu},     
	pdfsubject={hep-th},   
	pdfcreator={pdfLaTex},   
	pdfproducer={LaTex}, 
	pdfkeywords={},
	colorlinks=true
	,urlcolor=blue
	,anchorcolor=blue
	,citecolor=blue
	,filecolor=blue
	,linkcolor=blue
	,menucolor=blue
	,linktocpage=true
}

\allowdisplaybreaks[1] 

\crefname{figure}{Figure}{Figures}
\crefname{table}{Table}{Tables}
\crefname{definition}{Definition}{Definitions}
\crefname{proposition}{Proposition}{Propositions}
\crefname{claim}{Claim}{Claims}
\crefname{conjecture}{Conjecture}{Conjectures}

\def\beq{\begin{equation}}
	\def\eeq{\end{equation}}
\def\bc{\begin{cases}}
	\def\ec{\end{cases}}
\def\bal{\begin{aligned}}
	\def\eal{\end{aligned}}
\newcommand{\mc}{\mathcal}
\newcommand{\mf}{\mathfrak}

\arxivnumber{}
\title{Expanding the Weak Gravity Conjecture in AdS Space}
\author{Puxin Lin\,\orcidlink{0000-0003-0287-1662},}
\author{Alessandro Mininno\,\orcidlink{0000-0002-9593-0440},}
\author{Gary Shiu\,\orcidlink{0000-0003-1308-5202}}
\affiliation{Department of Physics, University of Wisconsin--Madison\\
	1150 University Avenue, Madison, WI 53706, USA}

\emailAdd{plin73@wisc.edu}
\emailAdd{mininno@physics.wisc.edu}
\emailAdd{shiu@physics.wisc.edu}

\abstract{
	In this work, we investigate the extension of our recent proposal for the Weak Gravity Conjecture (WGC) in AdS space to more general effective field theories. We first extend the conjecture to set-ups where moduli are present and we demand that a particle, produced during the decay of an extremal black hole via the Schwinger effect, is repelled close to the horizon of the black hole. 
	We interpret this condition as the universal criterion imposed by the WGC for any background. Interestingly, the constraint imposed by the WGC on the particle spectrum, 
	before the stabilization of the moduli, is independent of the background, resulting in the same bound both in Minkowski and in AdS space. For the case of AdS space, after the stabilization of the moduli, the WGC in AdS space is reproduced due to the formation of a non-singular horizon of the extremal black hole. In that case, the particle spectrum must satisfy the stronger of the two WGC conditions obtained before and after moduli stabilization.
	Finally, we use these results to argue that a similar version of the tower WGC should also apply to an AdS background. 
}

\begin{document}
	
	\maketitle
	
	\section{Introduction}
	\label{sec:introduction}

	One of the 
	central 
	conjectures in the Swampland program \cite{Vafa:2005ui} (see reviews \cite{Brennan:2017rbf,Palti:2019pca,vanBeest:2021lhn,Grana:2021zvf,Agmon:2022thq, Montero:2024qml}) is the Weak Gravity Conjecture (WGC) \cite{Arkani-Hamed:2006emk}.\footnote{We refer to specific reviews, e.g. \cite{Palti:2020mwc,Harlow:2022ich,Rudelius:2024mhq}, for in-depth discussions about the WGC.} Such a conjecture posits a constraint on the spectrum of charged massive particles within the Effective Field Theory (EFT), such that extremal black holes are able to decay. Many versions of the WGC have been proposed and refined.
	These WGCs represent, to date, some of the strongest
	quantum gravity criteria
	that distinguish EFTs belonging to the Landscape from those in the Swampland.

	In a recent work \cite{Lin:2025wfe}, a version of the WGC in AdS space has been proposed, based on the physical conditions under which an extremal Reissner--Nordstr\"om AdS (RN--AdS) black hole (BH) can decay via Schwinger effect \cite{Schwinger:1951nm}. The result was supported not only by the consistency with the Breitenlohner--Freedman (BF) instability condition \cite{Breitenlohner:1982bm,Gubser:2008px,Gubser:2008pf,Denef:2009tp} for the charged particle in the vicinity of the extremal black hole horizon, but also by the condition for a charged particle to be repelled by the black hole in the vicinity of its horizon. The BH-particle repulsive condition ensures that decay products can classically separate from the black hole, and resembles the Repulsive Force Conjecture (RFC) \cite{Heidenreich:2019zkl,Heidenreich:2020upe}, which is based on the idea that the gravitational force should be weaker than the gauge force.
	
	The results in \cite{Lin:2025wfe} more prominently showed that, to correctly formulate the WGC in backgrounds that are not flat, one should not simply impose the existence of a particle with charge-to-mass ratio exceeding that of an extremal black hole in that background. Instead, one would need to compute, using the extremal black hole solution, the actual condition that the particle can be produced and experiences a repulsive force near the black hole. Explicitly, in the case of a RN--AdS black hole, a charged particle must satisfy
	\begin{equation}\label{eq:WGCinAdS-intro}
		\frac{M_\PlD{D}^{D-2}g^2_\UoD{D}\mf{q}^2}{\mf{m}^2} \geq \gamma_\ttiny{D}(0)\underbrace{\frac{1+\frac{(D-1)(D-2)}{(D-3)^2}\eta^2}{1+\frac{D-1}{D-3}\eta^2}}_{\mathcal{R}(\eta)^2}\coma 
	\end{equation}
	where $\gamma_\ttiny{D}(0) = \frac{D-3}{D-2}$ is the extremality factor of an RN black hole in flat space and $\eta = \frac{r_h}{\ell_\AdSD{D}}$ is the ratio of the areal radius of the black hole and the AdS length. The condition set by \eqref{eq:WGCinAdS-intro} has the following implications:
	\begin{enumerate}
		\item First, \eqref{eq:WGCinAdS-intro} reduces to the WGC in flat space both in the limits of $\ell_\AdSD{D}\rightarrow \infty$ or $r_h\rightarrow 0$. The latter includes a special case of a singular AdS dilaton black hole with a vanishing horizon area. The AdS WGC bound derived from this solution is the same as in flat space.
		\item Second, the factor $\mathcal{R}(\eta)$ in \eqref{eq:WGCinAdS-intro} can be understood as the correction to the 
		flat space $\gamma_\ttiny{D}(0)$ factor at all orders in $\eta$ that involves the AdS contribution. 
	\end{enumerate}
	
	Using \eqref{eq:WGCinAdS-intro} as an anchor, we further the discussion of the AdS WGC to set-ups that contain moduli. In order for the charge-to-mass ratio of the particle to make sense at generic points in the moduli space, the way in which the particle mass and gauge coupling depend on the moduli must be related. In fact, as we will see, in the case of Einstein--Maxwell--Dilaton (EMD) theory, the mass of a probe particle and the gauge coupling will have the same dependence on the dilaton field. This is essential because the moduli can run to large values near the horizon of an extremal dilaton black hole. The extensions considered in this work aim
	to understand the consistency of the AdS WGC under Kaluza--Klein (KK) reduction. This is in part motivated by \cite{Heidenreich:2015nta}, where the authors sharpened the WGC in flat space by exploring theories related through dimensional reduction. However, a similar statement for the WGC is lacking in AdS space and the present work aims to fill in this gap.
	Having a formulation of the WGC in AdS that can be applied to EFTs obtained via dimensional reduction can be used as a starting point to test the  proposed criterion
	in top-down quantum gravity theories. 
	In fact, by demanding the WGC to be preserved under dimensional reduction,
	\cite{Heidenreich:2015nta} originally proposed the existence of infinitely many WGC-satisfying states that populate the full charge lattice. It was later understood that the WGC-satisfying states may furnish a sublattice \cite{Heidenreich:2016aqi,Montero:2016tif} or even just an infinite tower \cite{Andriolo:2018lvp}.
	More precisely,
	after dimensional reduction on a circle, one requires that the original $\U(1)$ and the additional KK $\U(1)$ satisfy the Convex Hull Condition \cite{Cheung:2014vva}, which is the extension of the WGC to multiple $\U(1)$ factors. 
	Under the assumption that every circle compactification radii is allowed, this condition is violated for arbitrarily small circle radii unless the original higher-dimensional theory contains an infinite number of super-extremal states, leading to the proposal of the tower Weak Gravity Conjecture (tower WGC), which postulates a tower of super-extremal states with arbitrarily large charge along any ray in the charge lattice of the original theory.
	
	\subsubsection*{Summary of the Results}
	
	In this paper, we formulate the WGC in AdS in the presence of moduli as the constraint on the particle spectrum that allows for an extremal black hole to decay. In the case of unstabilized moduli, the extremal black hole solution develops a singular horizon, rendering the horizon area zero. We showed that, by requiring the charged particle produced via Schwinger effect to be repelled by the black hole near its singular horizon, the WGC bound on the particle spectrum no longer probes the AdS scale and reduces to the WGC bound from dilaton black holes in flat space.  
	
	We next explore the stabilization of the moduli which leads to non-trivial modifications to the WGC bound. The stabilization allows for non-singular and finite area horizons for the charged AdS black holes, leading to a condition on the particle spectrum that is compatible with the proposal of the AdS WGC in \cite{Lin:2025wfe}. In particular, we conclude that whenever it is possible to construct non-singular extremal AdS black hole, competing effects from the moduli and the black hole size occur and one is tasked to identify the strongest condition as the AdS WGC.
	
	Finally, we use these results to argue that a similar version of the tower WGC should also apply to AdS backgrounds.

	\subsubsection*{Structure of the Paper}
	
	The paper is organized as follows. In Section \ref{sec:WGCasDecayBH}, we first review the WGC bound for both RN and EMD black holes in Minkowski space. We next obtain the WGC in AdS, recovering the result in \cite{Lin:2025wfe} and further extend it to the EMD--$\Lambda$ black hole in Section \ref{sec:EMDinAdSspace}. In Section \ref{sec:StabilizationDilaton}, we discuss the compatibility of the AdS WGC bound upon stabilization of the moduli. Unlike in flat space, stabilizing the moduli field no longer guarantees that the resulting extremal black hole can decay, and one would need to take the stronger condition before and after stabilization as the AdS WGC. In Section \ref{sec:TowardsTowerWGCAdS}, we show that the black hole decay prescription for the WGC in flat space requires a tower of states for consistency under dimensional reduction, and the same logic holds true to support a tower WGC in AdS space. In Section \ref{sec:conclusions}, we conclude and discuss possible future directions. We construct local solutions describing an EMD--$\Lambda$ black hole in Appendix \ref{app:EMDLambda}, and in Appendix \ref{app:tower} we report the explicit computation of the force relation between a particle and an EMD black hole carrying KK charges.
	
	\subsubsection*{Conventions and Notation}
	
	\begin{enumerate}
		\item We use a standard convention of $p$-form generalized symmetries in a $D$-dimensional EFT. A $p$-form global symmetry acts on dynamical objects with $(p+1)$-dimensional worldvolumes in spacetime. The symmetry is gauged by a $(p+1)$-form gauge field, with a $(p+2)$-form field strength. Moreover, we define
		\[|F_{p+2}|^2 \equiv F_{p+2}\wedge \star F_{p+2}=\frac{1}{(p+2)!}F_{\mu_1\ldots \mu_{p+2}}F^{\mu_1\ldots \mu_{p+2}}\fstop\]
		The gauge couplings associated to gauged $p$-form symmetries will be called $g_\UoD{p;D}$. We will focus mainly on gauged abelian $0$-form symmetries, in which case we will omit the label $p$ for $0$-form symmetries. 
		\item Unless otherwise specified, we work with $D$-dimensional EFTs compactified on a circle to $d=D-1$ dimensions.
		\item Capital calligraphic letters, such as $\mathcal{M}$ and $\mathcal{Q}$, will identify physical quantities associated to black holes. This is to distinguish them from scaled parameters, e.g. $M$ and $Q$, which sometimes enter black hole solutions. When the physical quantities are associated with particles, we will use $\mathfrak{fraktur}$ letters.
	\end{enumerate}
	
	\section{Weak Gravity Conjecture as Black Hole Decay Condition}
	\label{sec:WGCasDecayBH}
	
	In this section, we review results about the WGC relevant to our work. We first consider extremal black holes in Minkowski space, both with and without dilaton coupling. Next, we review the AdS WGC obtained in \cite{Lin:2024jug} and discuss its extension including moduli.
	
	\subsection{Reissner--Nordstr\"om Black Holes in Minkowski Space}
	\label{sec:RNinflatspace}
	
	We start with the simple case of an Einstein--Maxwell (EM) action in the Einstein frame: 
	\begin{equation}\label{eq:SEMacton}
		S \supset \frac{M_{\PlD{D}}^{D-2}}{2}\int_{\mathcal{M}_\ttiny{D}}  R\star\ID_\ttiny{D} -\frac{1}{2g_{\UoD{D}}^2}\int_{\mathcal{M}_\ttiny{D}} F_2\wedge \star F_2\coma
	\end{equation}
	where $M_\PlD{D}$ is the $D$-dimensional Planck mass, and $g_\UoD{D}$ is the $\U(1)$ gauge coupling.\footnote{In a $D$-dimensional theory, the gauge coupling associated to gauged $0$-form symmetries has mass dimensions $\frac{4-D}{2}$.} 
	
	This is the context in which the WGC was first formulated \cite{Arkani-Hamed:2006emk}, stating that, in order for extremal RN black holes to be able to decay, there must exist an object in the spectrum of the theory with mass $\fm$ and charge $\fq$ satisfying 
	\begin{equation}
		\label{eq:DdimWGC}
		\frac{M_\PlD{D}^{D-2} g_\UoD{D}^2\mf{q}^2}{\mf{m}^2}\geq \gamma_\ttiny{D}(0) \coma
	\end{equation}
	where we have defined
	\begin{equation}\label{eq:extremalityboundflatspace}
		\gamma_\ttiny{D}(0) = \frac{D-3}{D-2}\fstop
	\end{equation}
	In flat space, $\gamma_D(0)$ also corresponds to the charge-to-mass ratio of an extremal RN black hole, i.e., 
	\begin{equation} \label{eq:BH_extremality}
		\gamma_\ttiny{D}(0) \equiv \left.\frac{M_\PlD{D}^{D-2}g_\UoD{D}^2\mathcal{Q}^2}{\mathcal{M}^2}\right|_\ttiny{ext.}\coma
	\end{equation}
	where
	\begin{equation}
		\mathcal{Q} = \frac{1}{g_\UoD{D}^2}\int_{\mathcal{M}_{D-2}}\star F_2\coma
	\end{equation}
	and $\mathcal{M}$ is the ADM mass of the RN black hole.
	
	Because of \eqref{eq:BH_extremality}, the WGC in flat space is often phrased as the requirement for the existence of a super-extremal particle whose charge-to-mass ratio is larger than or equal to that of an extremal RN black hole. We emphasize that, while this phrasing is true in flat space, the use of the extremality bound of black holes can lead to confusion when the charge and mass of extremal black holes are not proportional. An example is extremal black holes in AdS space, whose extremality factor approaches zero for large black holes.
	
	The approach that is more faithful to the black hole decay argument is what has been proposed in \cite{Lin:2024jug}. By computing the Schwinger effect \cite{Schwinger:1951nm} of an extremal RN black hole, the authors discovered that charge production of extremal black holes cannot happen unless there exists a particle satisfying \eqref{eq:DdimWGC}, meaning that an RN black hole is, indeed, only able to decay via production of super-extremal particles.\footnote{The authors in \cite{Castellano:2025yur} found similar imaginary non-perturbative contributions to the effective action of BPS black holes that signifies the Schwinger effect in the presence of supersymmetry.}
	
	Another formulation of the WGC in flat space compares the strengths of gravitational and electromagnetic forces, stating that gravity is the weakest among the two. This is the proposal considered in \cite{Heidenreich:2019zkl} (building on \cite{Palti:2017elp}), which concluded that the WGC is equivalent to the requirement that a self-repulsive state exists (i.e., two copies of the same particle repel each other at long range). In \cite{Lin:2025wfe}, the repulsive force formulation was revisited. Instead of computing the force between two particles at long range, \cite{Lin:2025wfe} evaluated the force between a black hole and a particle near its horizon. From the particle-black hole force, it was found that repulsiveness near the horizon translates precisely to 
	\eqref{eq:DdimWGC}. There are two advantages of formulating the WGC as a repulsive condition between a black hole and a particle near the horizon. First, the condition can be thought of as the classical statement of black hole decay --- the particle must separate from the black hole upon creation. This classical picture and the quantum process (the Schwinger effect) complement each other and unify into the extremal black hole decay statement. Second, in the presence of a non-zero cosmological constant, the long range limit can break down. In AdS, for instance, massive objects experience a confining AdS potential and cannot separate to large distance. While in flat space, the different formulations lead to the same result, the black hole decay formulation makes the generalization to general spacetimes straightforward.

	\subsection{Einstein--Maxwell--Dilaton Black Holes in Minkowski Space}
	\label{sec:EMDinflatspace}
	
	The first generalization of the WGC for particles is obtained when one considers a dilaton field coupled to the $\U(1)$ gauge field, namely a theory given by the following EMD action:
	\begin{equation}\label{eq:SEMDaction}
		S \supset \frac{M_{\PlD{D}}^{D-2}}{2}\int_{\mathcal{M}_\ttiny{D}}  \left(R\star\ID_\ttiny{D} - \frac{1}{2} d \phi \wedge \star d \phi\right) -\frac{1}{2g_{\UoD{D}}^2}\int_{\mathcal{M}_\ttiny{D}} e^{-\alpha\phi} F_2\wedge \star F_2\coma
	\end{equation}
	where the electric charge under the $\U(1)$ above is defined as 
	\begin{equation}
		\mc{Q} = \frac{1}{g_\UoD{D}^2}\int_{\mathcal{M}_{D-2}}e^{-\alpha \phi}\star F_2\fstop
	\end{equation}
	
	The solution for EMD black holes has been considered, e.g., in \cite{Horowitz:1991cd}, and the implications to the WGC have been discussed in \cite{Heidenreich:2015nta}. While extremal solutions are singular at the horizon due to the presence of the dilaton, they can be considered as a limit of non-extremal EMD black holes. The charge and the ADM mass of the black hole can be expressed as functions of the inner and outer horizons $r_\pm$. Hence, considering the charge-to-mass ratio of general EMD black holes and taking the extremal limit $r_-\rightarrow r_+$, one concludes that
	\begin{equation}\label{eq:EMDextremalratio}
		\left.\frac{M_\PlD{D}^{D-2}g_\UoD{D}^2\mathcal{Q}^2}{\mathcal{M}^2}\right|_\ttiny{ext.} \equiv \gamma_\ttiny{D}(\alpha)  \coma
	\end{equation}
	where one identifies
	\begin{equation}\label{eq:extremalityboundflatspacealpha}
		\gamma_\ttiny{D}(\alpha) = \frac{D-3}{D-2}+\frac{\alpha^2}{2}\coma
	\end{equation}
	as the extremality factor for the EMD black hole. The black hole extremality formulation\footnote{Once again, we would like to stress that the extremality factor can be used to set the WGC bound only in Minkowski space. In general spacetime, one should recover the WGC bound from the decay condition of extremal black holes.} of the WGC would then instruct one to conclude that there must exist a particle satisfying 
	\begin{equation}
		\label{eq:DdimWGC-dilaton}
		\frac{M_\PlD{D}^{D-2}g_\UoD{D}^2\mf{q}^2}{\mf{m}^2}\geq \gamma_\ttiny{D}(\alpha) \fstop
	\end{equation}
	Equivalence between the black hole extremality and the black hole decay condition in flat space will be demonstrated in the later sections. The interesting fact about \eqref{eq:DdimWGC-dilaton} is that  \eqref{eq:extremalityboundflatspacealpha} is larger than \eqref{eq:extremalityboundflatspace}, which means that requiring the existence of a particle satisfying \eqref{eq:EMDextremalratio} for the whole moduli space given by the dilaton field is sufficient to guarantee that the WGC is satisfied after the stabilization of the dilaton. 
	
	\subsubsection{Force Condition for Singular EMD Black Holes} 
	
	In this section, we obtain the WGC bound for the EMD theory as the black hole decay condition. Extremal EMD black holes are singular at the horizon and no longer have a near horizon $AdS_2\times S^{D-2}$ geometry, which makes the explicit calculation of the Schwinger effect difficult. On the other hand, in \cite{Lin:2024jug,Lin:2025wfe}, the authors gave evidence that the decay condition is also set by requiring the charged particle produced via the Schwinger effect to be repelled close to the horizon of the extremal black hole.\footnote{The repulsive force condition for these black holes can be obtained by utilizing the explicit solution. Alternatively, one finds the same expression by following the analysis in Appendix \ref{app:EMDLambda} and setting $\Lambda_\AdSD{D}\rightarrow 0$. Note that the analysis in Appendix \ref{app:EMDLambda} can be applied even without having global solutions, which becomes useful in AdS space.} Since this computation will also be useful for Section \ref{sec:TowardsTowerWGCAdS}, we will briefly highlight the procedure and derive \eqref{eq:DdimWGC-dilaton} by computing the force on a charged particle in the EMD black hole background. 
	
	Making use of the results reported in \cite{Horowitz:1991cd,Heidenreich:2015nta}, we collect the solution to the equations of motion following from the action \eqref{eq:SEMDaction}:
	\begin{equation}
		\begin{split}
			ds^2 &= -f(r)^{\gamma_1}dt^2+f(r)^{\gamma_2-2}dr^2+r^2f(r)^{\gamma_2}d\Omega_{D-2}^2\coma\\
			F_2 &=\frac{g_\UoD{D}^2\mc{Q}}{\omega_{D-2}r^{D-2}}dt\wedge dr\coma\\
			e^{\alpha \phi} &= f(r)^{\frac{\alpha^2}{\gamma_\ttiny{D}(\alpha)}}\coma
		\end{split}
	\end{equation}
	where
	\begin{equation}
		f(r) = 1-\frac{r_h^{D-3}}{r^{D-3}}\coma \mc{Q}=\frac{(D-3)\omega_{D-2}M_\PlD{D}^\frac{D-2}{2}}{\sqrt{\gamma_\ttiny{D}(\alpha)}g_\UoD{D}}r_h^{D-3}
	\end{equation} 
	and we have also introduced
	\begin{equation}
		\gamma_1 = \frac{2(D-3)}{(D-2)\gamma_\ttiny{D}(\alpha)}\coma \gamma_2 = \frac{\alpha^2}{(D-3)\gamma_\ttiny{D}(\alpha)} \fstop
	\end{equation}
	We now consider the worldline of a charged particle with charge $\mf{q}$ and mass $\mf{m}(\phi)^2=e^{\alpha\phi}\mf{m}^2$,
	\beq
	S_p=\int d\tau \left(\mf{m}(\phi)\sqrt{-g_{\mu\nu}\dot{\xi}^\mu\dot{\xi}^\nu}+\mf{q}A_\mu \dot{\xi}^\mu\right)
	\eeq
	and we compute the radial force density that this particle feels close to the extremal horizon of the black hole,\footnote{The particle is taken to be momentarily stationary.} i.e. 
	\beq \label{eq:general_forceEMD}
	\begin{split}
		F&=g_{00}^{-1}\Gamma^1_{00}-g^{11}\frac{\partial_\phi\mf{m}(\phi)}{\mf{m}(\phi)}\phi'+\frac{\mf{q}}{\mf{m}(\phi)}\frac{g^{11}}{\sqrt{-g_{00}}}F_{01}\\
		&=g_{00}^{-1}\Gamma^1_{00}-\frac{\alpha}{2}g^{11}\phi'+\frac{\mf{q}}{\mf{m}}e^{-\frac{\alpha}{2}\phi}\frac{g^{11}}{\sqrt{-g_{00}}}A'_t\fstop
	\end{split}
	\eeq
	By requiring that close to the horizon $F(r)|_{r\rightarrow r_h}\geq 0$,\footnote{For EMD black holes, it is not necessary to require that the force is repulsive close to the horizon and the same result holds for long-range forces, like it has been shown in \cite{Heidenreich:2019zkl,Heidenreich:2020upe}.} we find that
	\begin{equation}\label{eq:reobtainWGCEMD}
		\frac{M_\PlD{D}^{D-2}g_\UoD{D}^2\mf{q}^2}{\mf{m}^2} \geq \gamma_\ttiny{D}(\alpha)\fstop
	\end{equation}
	We note that, since the dependence of the particle on the dilaton is the same as the gauge coupling in \eqref{eq:SEMDaction}, the charge-to-mass ratio\footnote{Here we have defined $g_\UoD{D}^2(\phi) = g_\UoD{D}^2e^{\alpha \phi}$.}
	\begin{equation}
		\mathfrak{z}^2 \equiv \frac{M_\PlD{D}^{D-2}g_\UoD{D}^2\mf{q}^2}{\gamma_\ttiny{D}(\alpha)\mf{m}^2}=\frac{M_\PlD{D}^{D-2}g_\UoD{D}^2(\phi)\mf{q}^2}{\gamma_\ttiny{D}(\alpha)\mf{m}(\phi)^2}\coma
	\end{equation}
	does not depend on the dilaton anymore, and it remains constant even if the dilaton value diverges at the singular horizon. The positivity of the force comes solely from the last contribution in \eqref{eq:general_forceEMD}, since the first two are negative. We conclude that \eqref{eq:reobtainWGCEMD}, or equivalently
	\begin{equation}
		\mathfrak{z}^2 \geq 1\coma
	\end{equation}
	is precisely the WGC in \eqref{eq:DdimWGC-dilaton}, as expected.
	
	\subsection{Reissner--Nordstr\"om Black Holes in AdS Space}
	\label{sec:RNinAdSspace}
	
	We move on to the results in \cite{Lin:2025wfe}, where the authors proposed a version of the WGC in AdS based on the requirement for an extremal RN--AdS black to be able to decay. Let us consider the following action in the Einstein frame
	\beq\label{eq:AdSEMDilaton}
	S \supset \frac{M_\PlD{D}^{D-2}}{2}\int_{\mathcal{M}_\ttiny{D}} \left(R -2\Lambda_\AdSD{D}\right)\star \ID_\ttiny{D}- \frac{1}{2g_\UoD{D}^2}\int_{\mathcal{M}_\ttiny{D}} F_2 \wedge \star F_2\,,
	\eeq
	where the cosmological constant is identified as $\Lambda_\AdSD{D}=-\frac{(D-1)(D-2)}{\ell^2_\AdSD{D}}$ with the AdS length scale $\ell_\AdSD{D}$. By considering the explicit solution for an extremal RN--AdS black hole, one finds
	\begin{equation}\label{eq:RNAdSextremalratio}
		\left.\frac{M_\PlD{D}^{D-2}g_\UoD{D}^2\mathcal{Q}^2}{\mathcal{M}^2}\right|_\ttiny{ext.} = \gamma_\ttiny{D}(0)  \frac{1+\frac{D-1}{D-3}\eta^2}{\left(1+\frac{D-2}{D-3}\eta^2\right)^2} \coma
	\end{equation}
	where $\eta = \frac{r_h}{\ell_\AdSD{D}}$, signaling the first and most important difference between the RN black hole in flat space and in AdS space. The charge-to-mass ratio for an extremal RN--AdS black hole is generally smaller than $1$ (in Planck units) and it goes to zero for large black holes, leading to the wrong conclusion that no constraint on the particle spectrum is necessary for large RN--AdS black holes to decay.
	
	However, as was shown explicitly in \cite{Lin:2025wfe}, merely requiring that the charge-to-mass ratio of a particle be greater than or equal to \eqref{eq:RNAdSextremalratio} is insufficient to ensure the decay of extremal RN--AdS black holes. In fact, computing the vacuum-vacuum amplitude associated with the Schwinger effect, one finds the following threshold value
	\begin{equation}
		\label{eq:WGCinAdS}
		\frac{M_\PlD{D}^{D-2}g^2_\UoD{D}\mf{q}^2}{\mf{m}^2} \geq \gamma_\ttiny{D}(0)\underbrace{\frac{1+\frac{(D-1)(D-2)}{(D-3)^2}\eta^2}{1+\frac{D-1}{D-3}\eta^2}}_{\mathcal{R}(\eta)^2}
		\coma
	\end{equation} 
	that asymptotes to
	\begin{equation}
		\label{eq:WGCinAdS-infinity}
		\frac{M_\PlD{D}^{D-2}g^2_\UoD{D}\mf{q}^2}{\mf{m}^2} \geq 1\fstop
	\end{equation} 
	This condition led to the formulation of the AdS WGC that clearly demonstrates the difference between the extremality bound and the WGC condition. Remarkably, the condition for a non-zero production rate of charged particles via the Schwinger effect is the same as that for a particle to be repelled by the RN--AdS black hole near the horizon \cite{Lin:2025wfe}. 
	
	It is clear that the WGC in AdS looks similar to \eqref{eq:DdimWGC}, but with a correction $\mathcal{R}(\eta)^2$ that depends on the ratio of the size of the RN--AdS black hole and the AdS length. This effect is not present in flat space where $\ell_\AdSD{D}\rightarrow \infty$. 
	
	\subsection{Einstein--Maxwell--Dilaton--\texorpdfstring{$\Lambda$}{} Black Holes in AdS Space}
	\label{sec:EMDinAdSspace}
	
	Let us now consider a modification to \eqref{eq:AdSEMDilaton} by adding a dilaton field $\phi$ coupled to the cosmological constant $\Lambda_\AdSD{D}$ and the field strength. The relevant EMD--$\Lambda$ action is
	\beq \label{eq:EMDLambda_action}
	\scalebox{0.94}{$\displaystyle
		S \supset \frac{M_\PlD{D}^{D-2}}{2}\int_{\mathcal{M}_\ttiny{D}} \left[\left(R -2\Lambda_\AdSD{D}e^{\beta\phi}\right)\star \ID_\ttiny{D}- \frac{1}{2} d \phi \wedge \star d \phi\right]- \frac{1}{2g_\UoD{D}^2}\int_{\mathcal{M}_\ttiny{D}} e^{-\alpha \phi}F_2 \wedge \star F_2
		\fstop$}
	\eeq
	As we noted in the previous section, it is insufficient to simply use the charge-to-mass ratio of the extremal solution as the WGC in AdS space --- we need to derive the black hole decay condition.
	While the Schwinger effect is difficult to compute due to a similar singularity of solutions also present in solutions of Section \ref{sec:EMDinflatspace}, we can still obtain the decay condition by computing the repulsive force condition near the horizon. Contrary to the flat space example, we are not aware of explicit solutions to our action \eqref{eq:EMDLambda_action}. Fortunately, the fact that the force is evaluated near the horizon provides us with some leverage --- it is sufficient to solve the equations of motion locally near the horizon, as we do in Appendix \ref{app:EMDLambda}. 
	
	With the local solution in Appendix \ref{app:singularEMDAdS}, we proceed to evaluate the force on the particle with charge $\mf{q}$ and mass $\mf{m}(\phi)^2=e^{\alpha\phi}\mf{m}^2$.\footnote{We note that the charge-to-mass ratio
		\begin{equation}
			\gamma_\ttiny{D}(\alpha)\mathfrak{z}^2 = \frac{M_\PlD{D}^{D-2}g_\UoD{D}^2\mf{q}^2}{\mf{m}^2}\coma
		\end{equation}
		is independent of the dilaton and in fact coincides with the one defined in flat space.} The explicit computation is performed in Appendix \ref{app:RFCforsingularEMDAdS}. We find that a repulsive force between the extremal black hole and the particle is only possible if the particle satisfies the following condition
	\begin{equation}
		\label{eq:DdimWGC-dilaton-AdS}
		\frac{M_\PlD{D}^{D-2}g_\UoD{D}^2\mf{q}^2}{\mf{m}^2}\geq \gamma_\ttiny{D}(\alpha) \coma
	\end{equation}
	which is the same condition one finds in flat space, i.e. \eqref{eq:DdimWGC-dilaton}. One should not find the coincidence surprising because the EMD--$\Lambda$ black holes are singular, having a vanishing horizon size, which makes the black hole incapable of probing the AdS effect when restricting only locally near its horizon. However, it is important to note that $\gamma_\ttiny{D}(\alpha)$ in AdS space does not play the role of the black hole extremality factor anymore, but it is a proportionality factor between the gravitational and electromagnetic forces. 
	
	\subsubsection{On the Stabilization of the Dilaton}
	\label{sec:StabilizationDilaton}
	
	One crucial point in Section \ref{sec:EMDinflatspace} was that requiring the WGC for the whole moduli space of the dilaton field was sufficient to guarantee that the WGC was satisfied after stabilizing the dilaton. This is because the flat space extremality factor $\gamma_\ttiny{D}(\alpha)$ is greater than $\gamma_\ttiny{D}(0)$. On the other hand, in \cref{sec:RNinAdSspace,sec:EMDinAdSspace}, the conditions for the WGC to be satisfied depend on the dilaton coupling $\alpha$ and the comparison of the RN--AdS black hole size compared to the AdS length, i.e. $\eta$. If 
	\begin{equation}
		\gamma_\ttiny{D}(\alpha) \geq \gamma_\ttiny{D}(0)\frac{1-\frac{(D-1)(D-2)}{(D-3)^2}\eta^2}{1-\frac{D-1}{D-3}\eta^2}\coma
	\end{equation}
	for all $\alpha$ and $\eta$, then requiring the WGC for an EMD--$\Lambda$ black hole is sufficient to guarantee the AdS WGC after stabilization of the dilaton. However, we see that such a condition is satisfied only if 
	\begin{equation}\label{eq:alphaconditionstability}
		\alpha^2 \geq \frac{2}{D-2} \frac{\frac{D-1}{D-3} \eta ^2}{1+\frac{D-1}{D-3}\eta ^2} \stackrel{\eta\rightarrow \infty}{\longrightarrow} \frac{2}{D-2} \coma
	\end{equation}
	meaning that depending on the coupling $\alpha$, requiring the WGC for the EMD--$\Lambda$ might not be sufficient to guarantee that the AdS WGC is satisfied once the dilaton is stabilized. This could have already been observed by considering the WGC in AdS for the largest RN--AdS black hole, i.e. \eqref{eq:WGCinAdS-infinity}, that leads to a condition that does not depend on the spacetime dimensions of the EFT, while the constraint imposed by \eqref{eq:DdimWGC-dilaton-AdS} has still an explicit dependence on them. 
	However, \eqref{eq:alphaconditionstability} is the value at which the excitations of the emergent string predicted in \cite{Lee:2018urn} asymptotes as well in the weak coupling limit.
	
	For general EFTs, without knowing the spectrum, one needs to impose the strongest condition of the two
	--- 
	with and without moduli stabilization ---
	to be sure that all black holes are able to decay. We will comment on \eqref{eq:alphaconditionstability} more below and in Section \ref{sec:conclusions}.
	
	The presence of a singular horizon is a consequence of having the gauge coupling being a function of the moduli, in our case the moduli is the dilaton field. If the moduli are stabilized, and the gauge coupling is fixed, the extremal black hole solution will have a finite horizon. In the following, we will construct a toy model for the stabilization of the dilaton to show that smooth horizons indeed form and the WGC reduces to the one proposed in \cite{Lin:2025wfe} and reviewed in Section \ref{sec:RNinAdSspace}.
	
	The dilaton field is expected to be stabilized in the low energy theory. As a toy model, we add to \eqref{eq:EMDLambda_action} a dilaton mass term, $\frac{1}{2}\mu^2\phi^2$, to stabilize it. Effectively, considering the equation of motion for the dilaton, it is subject to the following potential:
	\beq \label{eq:dilaton_potential}
	V(\phi)=\frac{1}{2}\mu^2\phi^2-2(-\Lambda_\AdSD{D}) e^{\beta\phi}-\frac{1}{R^{2(d-2)}}\frac{g_\UoD{D}^2\mc{Q}^2}{M_\PlD{D}^{D-2}\omega_{D-2}^2} e^{\alpha\phi}
	\fstop\eeq
	
\begin{figure}
		\centering
		\subfloat[]{\centering
			\begin{tikzpicture}
				\node (plot) {\includegraphics[width=0.49\textwidth,keepaspectratio]{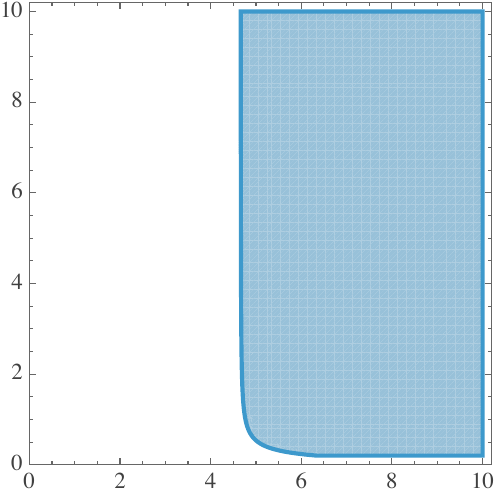}};
				\node (mu) [below=0mm of plot] {$\mu$};
				\node (phi) [left=0mm of plot] {$r_h$};
			\end{tikzpicture}
		}
		\caption{The parameter space for dilaton-stabilized black hole solutions. Horizontal axis represents the dilaton mass $\mu$ and the vertical axis represents the allowed black hole size $r_h$ that the dilaton can be stabilized, all in units of the AdS length $\ell_\AdSD{D}=1$. There is a minimum dilaton mass $\mu^2 \gtrsim|\Lambda_\AdSD{D}|$ requirement for the stabilization to happen. For any $\mu$, there is a minimum black hole size $r_\ttiny{h,min.}$ below which the curvature scale dominates over the dilaton mass and the black hole becomes singular.}
		\label{fig:dilaton_mass}
	\end{figure}

	When the spacetime is inhomogeneous, the dilaton cannot take on a global constant value since the effective coupling runs accordingly to the background (gravitational) source. However, near a smooth extremal horizon and at the asymptotical region, the dilaton can be viewed as a constant. Let $\phi_0$ be the dilaton value such that $V'(\phi_0)=0$ at the horizon. As should be noted, the potential is non-monotonic only when the dilaton is sufficiently heavy, namely, when the dilaton mass is larger than the scale set by the AdS scale $\Lambda_\AdSD{D}$ and the mass of the black hole. The general parameter space that allows for stabilization is shown in Figure \ref{fig:dilaton_mass}.
	
	The detailed calculation of the solution in the case of a stabilized dilaton is provided in Appendix \ref{app:stabilizedEMDAdS} and the computation of the force is done in Appendix \ref{app:RFCforstabilizedEMDAdS} and, eventually, leads to requiring a particle satisfying
	\beq\label{eq:generalRNAdSextremalratio}
	\frac{M_\PlD{D}^{D-2}g^2_\UoD{D} \mf{q}^2}{\mf{m}^2}\ge \gamma_\ttiny{D}(0)\frac{1+\left(\frac{D-1}{(D-3)^2}+\frac{D-1}{D-3}e^{\beta\phi_0}\right)\eta^2}{1+\frac{D-1}{D-3}\eta^2}\fstop
	\eeq
	The stabilization of the dilaton reduces $\gamma_\ttiny{D}(\alpha)$ in \eqref{eq:DdimWGC-dilaton-AdS} back to $\gamma_\ttiny{D}(0)$ originally found in \eqref{eq:RNAdSextremalratio}, but, at the same time, introduces a correction 
	\begin{equation}
		\tilde{\mc{R}}(\eta,\phi_0)^2 = \frac{1+\left(\frac{D-1}{(D-3)^2}+\frac{D-1}{D-3}e^{\beta\phi_0}\right)\eta^2}{1+\frac{D-1}{D-3}\eta^2}\coma 
	\end{equation} 
	that depends on the BH size. Such correction reduces to $\mc{R}(\eta)^2$ in \eqref{eq:RNAdSextremalratio} in the limit of $\phi_0\rightarrow 0$. As we sketched in \eqref{eq:alphaconditionstability}, to see if WGC is still satisfied upon stabilization, one needs to compare $\gamma_\ttiny{D}(\alpha)$ with $\gamma_\ttiny{D}(0)\tilde{\mc{R}}(\eta)^2$. For arbitrary values of the coupling $\alpha$, it is not always possible to guarantee that the WGC bound in \eqref{eq:DdimWGC-dilaton-AdS} is sufficient to guarantee the WGC in \eqref{eq:generalRNAdSextremalratio}. This means that, in general, satisfying the WGC in AdS for every value of the moduli does not imply the WGC in AdS after the naked singularities are resolved. This is one key difference from the WGC known in the literature formulated for asymptotically flat space. Interestingly, the critical value of $\alpha$ found in \eqref{eq:alphaconditionstability}, i.e.
	\begin{equation}
		\frac{\alpha^2}{2} = \frac{1}{D-2}\coma
	\end{equation}
	resembles the scaling behavior of the mass of the lighest possible tower of particles parametrizing the infinite distance limits in moduli space \cite{Etheredge:2022opl}, as predicted by the Distance Conjecture \cite{Ooguri:2006in}. In \cite{Etheredge:2022opl}, it has been argued that every infinite-distance limit in moduli space features at least one tower which satisfies the Distance Conjecture with scaling equal to \eqref{eq:alphaconditionstability}. Under the assumption that the conclusions also apply  to asymptotically AdS spaces, these towers will also satisfy the WGC in AdS even after stabilization of the moduli. We provide more comments about possible future directions in Section \ref{sec:conclusions}.
	
	\section{Motivations for a Tower Weak Gravity Conjecture in AdS Space}
	\label{sec:TowardsTowerWGCAdS}
	
	We turn to a final extension: the tower WGC in AdS space. Taking a step back, we first show how the black hole decay argument leads to the existence of a tower of states in flat space, and then comment on why the same argument works when there is a negative cosmological constant. In particular, we will consider the dimensional reduction of a $D$-dimensional EMD black hole, following the convention in \cite{Heidenreich:2015nta}.\footnote{Note that there are some changes to the notion used.} Starting from the action in \eqref{eq:SEMDaction}, we reduce to $d=D-1$ dimensions with the ansatz:
	\begin{equation}
		\begin{split}
			ds_\ttiny{D}^2 &= e^{\frac{\lambda}{d-2}}ds_d^2+e^{-\lambda}(dy+r_{S^1}B_1)^2\coma \\
			F_2 &= \tilde{F}_2 + \frac{1}{2\pi}F_1 \wedge \left(\frac{dy}{r_{S^1}}+B_1\right)\coma \\
			A_1 &= \tilde{A}_1+\frac{1}{2\pi}A_0\wedge \left(\frac{dy}{r_{S^1}}+B_1\right)\coma
		\end{split}
	\end{equation}
	where $A_0$ is a compact axion with field strength $F_1 = dA_0$. In the following, we will ignore this term because the black hole will not be charged under this. However, the field strength $\tilde{F}_2$ satisfies the following Bianchi identity:
	\begin{equation}
		d \tilde{F}_2 = \frac{1}{2\pi}H_2\wedge F_1\fstop
	\end{equation}
	The dimensionally reduced action is 
	\begin{equation} \label{eq:EMDreducedaction}
		\scalebox{1}{$
			\begin{aligned}
				S \supset & \, \frac{M_\PlD{d}^{d-2}}{2}\int_{\mathcal{M}_d} \left[R\star \ID_d -\frac{d-1}{4(d-2)}d\lambda \wedge \star d\lambda-\frac{1}{2e_\KK^2}e^{-\frac{d-1}{d-2}\lambda}|H_2|^2-\frac{1}{2}d\phi \wedge \star d\phi\right]+\\
				&-\int_{\mathcal{M}_d}\left(\frac{1}{2g_\UoD{d}^2} e^{-\alpha \phi -\frac{1}{d-2}\lambda}|\tilde{F}_2|^2 
				\right)\coma
			\end{aligned}$}
	\end{equation}
	where recall that $M_\PlD{d}^{d-2} = (2\pi r_{S^1})M_\PlD{D}^{D-2}$, while
	\begin{equation}\label{eq:eKKcoupling}
		\frac{1}{e_\KK^2}= \frac{1}{2}r_{S^1}^2M_\PlD{d}^{d-2} \coma
	\end{equation}
	and we have already turned off the axion field. 
	
	Before proceeding, let us pause to comment on the choice of starting from an EMD black hole. Starting from a EMD black hole with the dilaton being a modulus, or with a RN black hole does not change the computation that follows because we will keep $\gamma_\ttiny{D}(\alpha)$ to be the extremality factor in the original $D$-dimensional theory. As shown in \cite{Heidenreich:2015nta}, in the presence of a massless dilaton, the extremality factor is preserved under dimensional reduction, otherwise, it will be $\gamma_\ttiny{D}(0)> \gamma_\ttiny{D$-$1}(0)$. In the case of the WGC in AdS in Section \ref{sec:RNinAdSspace}, if the dilaton is massless, then $\gamma_\ttiny{D}(\alpha)=\gamma_\ttiny{D$-$1}(\alpha)$, but, even for massive dilaton, the WGC asymptotically approaches \eqref{eq:WGCinAdS-infinity}, which is independent of the spacetime dimensions. 
	
	In particular, by looking at the coupling of the radion $\lambda$ to the $F_2$ field strength in \eqref{eq:EMDreducedaction}, we notice that even when starting from a particle satisfying \eqref{eq:WGCinAdS-infinity} in $D$ dimensions, it will satisfy the WGC with massless radion in $d=D-1$ dimensions, since the coupling is larger than \eqref{eq:alphaconditionstability}. 
	
	We can now proceed in computing the extremality factor $\gamma_\KK$ for a black hole charged purely under a KK $\U(1)$ in two ways. One way is to realize that KK particles of massless particles satisfy
	\begin{equation}
		\mf{m}_\KK(\lambda)^2 = e^{\frac{d-1}{d-2}\lambda}\frac{\mf{q}_\KK^2}{r_{S^1}^2}\coma
	\end{equation}
	so that, using \eqref{eq:eKKcoupling} 
	\begin{equation}
		\frac{e_\KK^2e^{\frac{d-1}{d-2}\lambda} M_\PlD{d}^{d-2} \mf{q}^2_\KK}{\mf{m}_\KK(\lambda)^2} \equiv \gamma_\KK = 2\fstop
	\end{equation}
	This computation was possible because we are assuming that the particle couples to the moduli in the same way as the black hole does. The other possibility is to interpret the KK black hole with radion $\lambda$ as an EMD black hole, with canonically normalized dilaton
	\begin{equation}
		\hat{\lambda} = \sqrt{\frac{d-1}{2(d-2)}}\lambda\fstop
	\end{equation}
	The coupling of the dilaton $\hat{\lambda}$ with the field strength $H_2$ produces an effective coupling 
	\begin{equation}
		\frac{\alpha_\KK^2}{2} = \frac{d-1}{d-2}\coma
	\end{equation}
	leading to $\gamma_\KK = 2$ by using \eqref{eq:extremalityboundflatspacealpha}.\footnote{This computation is also interesting to conclude that BPS states such as KK modes simply satisfy the WGC but they do not saturate it after the stabilization of the radion. It is only when the radion is massless that the only KK modes that satisfy the WGC are those of massless particles that saturate it. This remains true for the WGC in AdS as we have seen in Section \ref{sec:StabilizationDilaton}. In general, supersymmetry leaves radions or, possibly present, dilatons massless in the infrared, and the only objects that saturate the WGC both in flat or AdS space would be BPS states.} In the following, it will not be important to work with canonically normalized radion since we will mostly interested in the charge-to-mass ratio of the $D$-dimensional particle. However, it is possible to work with $\hat{\lambda}$ by rescaling the field in the solution accordingly.
	
	The black hole solution is constructed by considering first an extremal electrically charged black hole in $d=D-1$ dimensions under only $\tilde{F}_2$, i.e.
	\begin{equation}
		\begin{split}
			ds^2 &= -f(r)^{\gamma_1}dt^2+f(r)^{\gamma_2-2}dr^2+r^2f(r)^{\gamma_2}d\Omega_{d-2}^2\coma\\
			\tilde{F}_2 &= \frac{g^2_\UoD{d} \mc{Q}}{\omega_{d-2}r^{d-2}}dt\wedge dr\coma\\
			e^{\lambda} &= f(r)^{\frac{2}{(d-1)\gamma_\ttiny{D}(\alpha)}}\coma\\
			e^{\alpha \phi} &= f(r)^{\frac{\alpha^2}{\gamma_\ttiny{D}(\alpha)}}\coma
		\end{split}
	\end{equation}
	where we have defined $f(r) = 1-\dfrac{r_h^{d-3}}{r^{d-3}}\in (0,1)$ and 
	\begin{equation}
		\gamma_1 = \frac{2(d-3)}{(d-2)\gamma_\ttiny{D}(\alpha)}\coma \gamma_2 = \frac{2}{d-3}-\frac{2}{(d-2)\gamma_\ttiny{D}(\alpha)} \fstop
	\end{equation}
	Then, to generate a KK charge, we perform a Lorentz boost along the $D-$th dimension,
	\begin{equation}
		t \rightarrow ut+v y\coma y \rightarrow uy+vt\coma
	\end{equation}
	with $u = \sqrt{v^2+1}$, such that the radion becomes
	\begin{equation}
		e^{-\lambda'}  = e^{-\lambda}\left(u^2-v^2f(r)^{\frac{2}{\gamma_\ttiny{D}(\alpha)}}\right) = e^{-\lambda}f_{\lambda}(r)\coma
	\end{equation}
	and the fully boosted solution is
	\begin{equation}\label{eq:boostedEDMsol}
		\begin{split}
			ds^2_d &= -f(r)^{\gamma_1}f_{\lambda}(r)^{-\frac{d-3}{d-2}}dt^2+f(r)^{\gamma_2-2}f_{\lambda}(r)^{\frac{1}{d-2}}dr^2+r^2f(r)^{\gamma_2}f_{\lambda}(r)^{\frac{1}{d-2}}d\Omega_{d-2}^2\coma\\
			\tilde{F}_2 &= \frac{g_\UoD{d} M_\PlD{d}^\frac{d-2}{2}u\mc{Q}}{r^{d-2}f_{\lambda}(r)}dt\wedge dr\coma\\
			H_2 &= \frac{u}{vr_{S^1}}dt\wedge df_{\lambda}(r)^{-1}\coma \\
			e^{\lambda} &= f(r)^{\frac{2}{(d-1)\gamma_\ttiny{D}(\alpha)}}f_{\lambda}(r)^{-1}\coma\\
			e^{\alpha \phi} &= f(r)^{\frac{\alpha^2}{\gamma_\ttiny{D}(\alpha)}}\fstop
		\end{split}
	\end{equation}
	
	The aim is to consider the force density on a particle charged under the original and the KK $\U(1)$s, with charges $\mf{q}$ and $\mf{q}_\KK$ respectively. The mass of such a KK mode will also depend on the moduli $\phi$ and $\lambda$. More precisely, the mass is given by\footnote{The mass is found by performing accordingly the KK reduction of a $D$-dimensional scalar field and identifying its mass in the $d$-dimensional theory.}
	\begin{equation}
		\label{eq:scalar_mass}\fm(\phi,\lambda)^2=e^{\frac{\lambda}{d-2}}\left(e^{\alpha\phi}\fm^2+e^{\lambda}\frac{\mathfrak{q}_\KK^2}{r_{S^1}^2}\right)\fstop
	\end{equation}
	The assumption is that we started from a $D$-dimensional EFT that was satisfying the WGC in $D$ dimensions, i.e. there exists a particle such that
	\begin{equation}
		\mf{z}_\ttiny{D}^2 \equiv \frac{M_\PlD{D}^{D-2}g_\UoD{D}^2\mf{q}^2}{\gamma_\ttiny{D}(\alpha)\mf{m}^2}\geq 1\fstop
	\end{equation}
	The strategy is to isolate $\mf{z}_\ttiny{D}$ in the expression of the force density and check if the existence of only one particle is sufficient to satisfy the WGC in the $d$-dimensional theory for all choice of the circle radius $r_{S^1}$. We will eventually see that just having one particle satisfying the WGC in $D$-dimension is not enough and we actually need a whole tower of such particles. In practice, we consider the force density in the black hole background of \eqref{eq:boostedEDMsol} of a KK-particle with charge-to-mass vector
	\begin{equation}\label{eq:ctmratioforKKparticle}
		\begin{split}
			\vec{\mf{z}} &=\frac{1}{\mf{m}(\phi,\lambda)}\left( \frac{M_\PlD{d}^{\frac{d-2}{2}}e^{\frac{\alpha}{2}\phi+\frac{\lambda}{2(d-2)}}g_\UoD{d}\mf{q}}{\sqrt{\gamma_\ttiny{D}(\alpha)}},\frac{M_\PlD{d}^{\frac{d-2}{2}}e^{\frac{d-1}{2(d-2)}\lambda}e_{\KK}\mf{q}_\KK}{\sqrt{\gamma_{\KK}}} \right) \\
			&= \frac{1}{\sqrt{e^{\alpha\phi}\fm^2+e^{\lambda}\frac{\mathfrak{q}_\KK^2}{r_{S^1}^2}}}\left( \frac{M_\PlD{d}^{\frac{d-2}{2}}e^{\frac{\alpha}{2}\phi}g_\UoD{d}\mf{q}}{\sqrt{\gamma_\ttiny{D}(\alpha)}},e^{\frac{1}{2}\lambda}\frac{\mf{q}_\KK}{r_{S^1}} \right) \\
			&= \frac{1}{\sqrt{e^{\alpha\phi}\fm^2 r_{S^1}^2+e^{\lambda}\mathfrak{q}_\KK^2}}\left( r_{S^1}e^{\frac{\alpha}{2}\phi}\fm \mf{z}_\ttiny{D},e^{\frac{1}{2}\lambda}\mf{q}_\KK \right)\\
			& = \frac{1}{\sqrt{1+e^{\lambda-\alpha\phi}\frac{\mathfrak{q}_\KK^2}{\fm^2 r_{S^1}^2}}}\left(\mf{z}_\ttiny{D},e^{\frac{1}{2}\lambda-\frac{\alpha}{2}\phi}\frac{\mf{q}_\KK}{\fm r_{S^1}} \right)\fstop
		\end{split}
	\end{equation}
	
	Each component of the charge-to-mass ratio in \eqref{eq:ctmratioforKKparticle} will contribute to the Coulomb force for each $\U(1)$ respectively. We point out a difference between $\mf{z}_\ttiny{D}$ and $\vec{\mf{z}}$: the former does not depend on the specific value of the dilaton $\phi$, since it has been defined in the original $D$-dimensional theory as the condition that the probe particle must satisfy to allow the decay of the $D$-dimensional EMD black hole. The latter, instead, will change its value depending on the values of the moduli $(\phi,\lambda)$. This is somewhat expected, because we are allowing the gauge couplings to effectively depend on the moduli, and $e^{\frac{1}{2}\lambda-\frac{\alpha}{2}\phi}$ measures the difference in the running of couplings between the KK $\U(1)$ and the dimensionally reduced original $\U(1)$. 
	In particular, we need to require 
	\begin{equation}\label{eq:towercondition}
		\frac{\mf{z}_\ttiny{D}^2}{1+e^{\lambda-\alpha\phi}\frac{\mathfrak{q}_\KK^2}{\fm^2 r_{S^1}^2}} \geq C^2 \Longrightarrow \fm^2 r_{S^1}^2\geq \frac{C^2\mathfrak{q}_\KK^2}{\mf{z}_\ttiny{D}^2-C^2}e^{\lambda-\alpha\phi}\coma
	\end{equation}
	for some constant $C$. As we compute in Appendix \ref{app:tower}, $C=\frac{1}{\gamma_\ttiny{D}(\alpha)}$ and the requirement for the force to be repulsive is
	\beq \label{eq:rcondition-maintext}
	\mf{m}^2 r_{S^1}^2\ge \frac{\mathfrak{q}_{\KK}^2e^{\lambda-\alpha\phi}}{\gamma_\ttiny{D}(\alpha)^2\mf{z}_\ttiny{D}^2-1}\coma
	\eeq
	The contribution of $e^{\lambda-\alpha\phi}$ is making the expression diverging at the horizon, since both $e^\lambda$ and $e^{\alpha \phi}$ go to zero. 
	However, this issue can be resolved by moving away from the horizon\footnote{If one instead would like to evaluate the long range force and take $r\rightarrow\infty$, the scalar dependent factor becomes $1$.} such that their values are finite and the EFT can be trusted. The direction of the inequality is not modified, and the requirement on $\mf{z}_\ttiny{D}$ for any fixed choice of $r_{S^1}$ is still required. This argument is similar to that which made use of the convex hull condition in \cite{Heidenreich:2015nta}, and to the argument in support of a strong form of the RFC in \cite{Heidenreich:2019zkl}.  Hence, we have shown that even using the RFC between a black hole and a probe particle, it is still possible to argue that for any $\mf{z}_\ttiny{D}$ there exists a minimal radius $r_{S^1}^{\ttiny{min.}}$ below which the force becomes attractive near the horizon of the black hole. Under the assumption that all possible radii of compactifications are allowed, we need a tower of states if we want to allow a charged KK black hole to decay, in the same spirit for which it has been argued for a tower in previous papers (see, e.g.,
	\cite{Heidenreich:2015nta,Montero:2016tif,Andriolo:2018lvp,FierroCota:2023bsp}). 
	
	We can easily extend the argument for a tower of states into AdS space. Note that, similar to the single U(1) dilaton black hole, the extremal solution \eqref{eq:boostedEDMsol} has zero horizon size. Therefore, the flat space solution is a good approximation of the AdS counterpart near the horizon and the force on the particle for both cases is the same to leading order in distance to the horizon. Besides, the AdS contribution to the force between any two massive objects is negative, meaning that the net force on the particle near the black hole horizon in AdS can only be more attractive than when $\Lambda_\AdSD{d}=0$. We can then safely conclude that \eqref{eq:towercondition}, derived near the black hole horizon, continues to hold with the caveat that the constant $C$ receives parametrically small corrections. Moreover, since we are considering gauge couplings that depends on the two moduli, the requirement on the particle before the KK-reduction imposed by the WGC is the same as in flat space.

	\section{Discussion and Conclusions}
	\label{sec:conclusions}
	
	The main goal of this paper is twofold: to clarify what the physical requirements are that lead to the WGC bound and to extend the discussion of the AdS WGC in \cite{Lin:2025wfe} to include moduli. We argue that, while in flat space the black hole extremality, long range force and the black hole decay formulations lead to the same WGC bound, the latter has a closer connection to the original proposal of the WGC and resolves ambiguities that can arise when formulating the WGC in general spacetimes. We make use of the black hole decay formulation, taking into account the Schwinger effect of the black hole and the separation of the produced particle from the extremal black hole, to obtain the WGC in AdS in the presence of moduli. We further discuss two scenarios relevant to the derived AdS WGC: stabilization of the moduli and dimensional reduction. We find that, unlike in flat space, one has to identify the stronger condition on the charged spectrum, before and after stabilization, as the AdS WGC associated with the EFT. Regarding dimensional reduction, we find indications of the existence of an infinite tower of states in flat space, in agreement with the existing literature, and in AdS space. 
	
	\subsubsection*{Outlooks}
	
	In this work, we have found that whenever there is a moduli space associated to unstabilized fields near the black hole horizon, the WGC in AdS reduces formally to the expression of the WGC known in flat space. This means that to see the effects of the cosmological constant on the WGC bound, the extremal black hole, as a solution to the EFT in AdS, must have a non-vanishing horizon size.
	
	A direction of great importance and significance is to test the proposals in explicit string theory or M-theory compactifications. While this work has an aim to improve our understanding of the WGC in AdS in the presence of moduli and in case of circle compactifications, we have not computed explicitly the Schwinger production rate of black holes in the case of an $\AdS_{D-1}\times S^1$ background nor did we obtain our AdS EFT directly from string compactifications. It would be interesting to extend our findings to $\AdS_m\times S^n$ compactifications, and understand if the relations among the AdS scale, the KK scale and the black hole horizon have some noticeable impacts on the WGC bound. If achieved, this would finally clarify and streamline the difference and the compatibility of our proposal of the WGC in AdS and the ones in the literature \cite{Nakayama:2015hga,Aharony:2021mpc,Cho:2025xod}.
	
	Relatedly, progress in the previously mentioned direction will facilitate further tests of the tower WGC in AdS with the proposal of the Minimal WGC \cite{FierroCota:2023bsp}. In that work, the authors considered F-/M-theory compactifications on Calabi--Yau (CY) threefolds, noting that the only towers of states that are required in order for the WGC to be preserved under dimensional reduction are those predicted by the Emergent String Conjecture (ESC) \cite{Lee:2019wij}. This means that there is usually a natural definition of minimal radius beyond which one cannot consider the EFT as the result of a circle compactification of a higher-dimensional EFT, and such a radius is set whenever the mass of the KK tower is comparable with the mass of the minimal black hole of the EFT. It would be interesting to check if similar reasoning can be used for the tower WGC in AdS, where, by construction, we require that the state repelled by the extremal black hole is a particle state. 
	
	Another observation is about scale separated vacua in AdS. For the case of unstabilized dilaton or radion, our findings are that the requirement of the WGC on the particle spectrum is the same as in flat space. For example, in the presence of supersymmetry, all protected operators saturate the WGC as long as the moduli remain massless. If supersymmetry prevents the dilaton and the radion to acquire a mass, then the WGC would be satisfied by these states and the theory will be compatible with Quantum Gravity. Interestingly, if we assume that the dilaton or radion acquires a mass in the infrared, the WGC in AdS reduces to the one proposed in \cite{Lin:2025wfe}, and the objects that previously saturated the WGC now satisfy it if their dependence on the moduli was at least the one predicted by the distance conjecture \cite{Etheredge:2022opl}. This means that, if through some mechanism, it was possible to realize a scale-separated vacuum in AdS, the WGC in AdS proposed in \cite{Lin:2025wfe} is still satisfied by the towers of states that were present before the stabilization of the moduli. So, if the corrections to the charge-to-mass ratio of these states are small enough to keep it larger than $1$ (independent of the spacetime dimension), the EFT would still be compatible with the WGC in AdS. On the other hand, the argument for a tower WGC in AdS we have proposed in this paper is based on requiring the Coulomb force to be the strongest force so that it overcomes all the other attractive forces. Under the assumption that every possible circle radius compactification is allowed, a tower of particles in the original theory is necessary in order for the Coulomb force to be the most repulsive force for any choice of the radius. So, it would be interesting to understand how this assumption reconcile with the AdS Distance Conjecture \cite{Lust:2019zwm} that prevents an arbitrarily light KK scale relative to the AdS scale. Assuming that such conjecture provides for a natural ``minimal radius" for the internal space of compactification, the tower WGC in AdS might not always be required, but it is only required for those towers predicted by the ESC, as it was concluded in \cite{FierroCota:2023bsp} for Minkowski vacua. It would be then interesting to refine the argument for the WGC in AdS to check if there is an explicit interplay between the requirement of a tower to satisfy the WGC in AdS and the presence/absence of scale separation in AdS compactifications of string theory. This, once again, will require understanding the relation between the charged towers of states required by the tower WGC and those parametrizing infinite distance limits in AdS. We plan to address this relation in the future.

	One final interesting observation is that in \cite{Heidenreich:2015nta}, a way to obtain the axion WGC was by the inclusion of the coupling of the dilaton that is modifying the extremality bound for gauged $p$-form symmetries. In this way, it has been possible to extend the WGC to gauged $(-1)$-form symmetries. Our result shows that, in the presence of moduli, the formulation of the WGC in AdS reduces to the usual formulation in flat space, hinting that the form of the axion WGC would also look similar in AdS space, despite the different asymptotics. However, this observation is only based on the results obtained when applying the repulsive force conjecture to a system of a black hole and a probe particle. The fact that there is no notion of force or decay for instantons has been an obstacle for formulating precisely the axion WGC, although there is evidence that the axion WGC bound is set by Euclidean wormholes \cite{Andriolo:2020lul, Andriolo:2022rxc}. Another obstacle in formulating the axion WGC is that unlike for black holes, there is a lack of explicit (non-singular) string theory constructions to test the conjectured criteria (see \cite{Loges:2023ypl} for recent effort). 
	It would be then interesting to understand how one can extend the proposal of the WGC in AdS to gauged generalized symmetries \cite{Gaiotto:2014kfa}.

	\acknowledgments
	
	The authors thank Rafael \'Alvarez-Garc\'ia, Ivano Basile, Amineh Mohseni, Vinicius Nevoa, Cumrun Vafa and Matteo Zatti for helpful discussions and comments. A. M. thanks Harvard University and the Harvard Swampland Initiative for the hospitality during the completion of this work. This work is supported in part by the DOE grant DE-SC0017647.
	
	\appendix
	
	\section{Einstein--Maxwell--Dilaton--\texorpdfstring{$\Lambda$}{Lambda} Black Hole Solutions in AdS Space}
	\label{app:EMDLambda}
	
	In this appendix, we seek electrically charged solutions to the EMD--$\Lambda$ action \eqref{eq:EMDLambda_action}. For solutions that are spherically symmetric, the metric can be cast into the form
	\beq
	ds^2=-U(r)dt^2+\frac{dr^2}{U(r)}+R^2(r)d\Omega_{D-2}^2\fstop
	\eeq
	The Maxwell equation can be integrated and expressed in terms of the metric as
	\beq
	F^{01}=\frac{g_\UoD{D}^2\mc{Q}}{\omega_{D-2}}\frac{e^{\alpha\phi}}{R^{D-2}}\coma
	\eeq
	and the dilaton field satisfies the following equation
	\beq
	\frac{1}{R^{D-2}}\frac{d}{dr}\left(R^{D-2}U(r)\phi'(r)\right)=-\frac{\de V(\phi)}{\de \phi}\coma
	\eeq
	where the dilaton potential is given by
	\beq \label{eq:dilaton_potential_no_mass}
	V(\phi)=-2(-\Lambda_\AdSD{D})e^{\beta\phi}-\frac{g_\UoD{D}^2\mc{Q}^2}{M_\PlD{D}^{D-2}\omega_{D-2}^2}\frac{ e^{\alpha_\ttiny{D}\phi}}{R^{2(D-2)}}\coma
	\eeq
	In writing the equations of motion, we denote $(\cdot)' = \frac{d}{dr} (\cdot)$. The system of independent equations of motion is collected as follows
	\beq \label{eq:field_eqns}
	\scalebox{0.82}{$
		\renewcommand{\arraystretch}{1.7}
		\displaystyle\left.
		\begin{array}{rcl}
			\frac{R''}{R}+\frac{(\phi')^2}{2(D-2)} & = & 0\coma \\
			\frac{1}{R^{D-2}}\frac{d}{dr}\left(R^{D-2}U\phi'\right)
			+2\beta \Lambda_\AdSD{D} e^{\beta\phi} & = & \frac{\alpha}{R^{2(D-2)}}  \frac{g_\UoD{D}^2\mc{Q}^2}{M_\PlD{D}^{D-2}\omega_{D-2}^2} e^{\alpha\phi}\coma \\
			\frac{1}{2}U''+\frac{D-2}{2}U'\frac{R'}{R}+\frac{2}{D-2}\Lambda_\AdSD{D} e^{\beta\phi} & = & \frac{D-3}{D-2}\frac{1}{R^{2(D-2)}} \frac{g_\UoD{D}^2\mc{Q}^2}{M_\PlD{D}^{D-2}\omega_{D-2}^2} e^{\alpha\phi}\coma\\
			-U''+\frac{(D-2)(D-3)}{R^2}-\frac{2(D-2)}{R}\left(UR'\right)'-\frac{(D-2)(D-3)}{R^2}U(R')^2-\frac{2D}{D-2}\Lambda_\AdSD{D} & = & \frac{D-4}{D-2}\frac{1}{R^{2(D-2)}} \frac{g_\UoD{D}^2\mc{Q}^2}{M_\PlD{D}^{D-2}\omega_{D-2}^2} e^{\alpha\phi}\fstop
		\end{array}\right.$}
	\eeq
	
	In the next sections, we solve the above equations locally near the extremal horizon for two situations: when the dilaton potential does not admit a minima and when the dilaton can be stabilized at the potential minima.
	
	\subsection{Dynamical Dilaton and Singular Extremal Solutions}
	\label{app:singularEMDAdS}
	
	In \eqref{eq:dilaton_potential_no_mass}, it is clear that we have a run-away potential without minima. In this case, the charged dilatonic black hole solutions have singularities at the inner horizon of the black hole where the dilaton diverges. When the extremal limit is taken, the singularity is pushed out towards the black hole horizon, forcing the extremal horizon to become zero-sized. The divergence of the dilaton field is logarithmic and the extremal solution can be expanded near the horizon as
	\beq \label{eq:local_expansion_singular}\renewcommand{\arraystretch}{1.7}\left.
	\begin{array}{rcl}
		\phi(r) &=& A\ln{\rho}+\phi_0+\cdots\coma\\
		R(r)&=&R_0\rho^\lambda+\cdots\coma\\
		U(r)&=&U_0\rho^\delta+\cdots\coma
	\end{array}\right.\eeq
	where $\rho=\frac{r-r_h}{\ell_0}$, $\ell_0$ being some length scale. With this expansion, the field equations \eqref{eq:field_eqns} reduce from differential equations to algebraic ones, i.e.
	\beq \label{eq:singular_equations}
	\scalebox{0.93}{$\displaystyle\renewcommand{\arraystretch}{1.7}\left.
		\begin{array}{rcl}
			\lambda &=&\frac{1}{2}\left(1\pm\sqrt{1-\frac{2A^2}{D-2}}\right)\coma\\
			\delta&=&2+\alpha A-2(D-2)\lambda\coma\\
			U_0&=& \frac{1}{A R_0^{2(D-2)}[\delta+(D-2)\lambda-1]} \alpha \frac{g_\UoD{D}^2\mc{Q}^2}{M_\PlD{D}^{D-2}\omega_{D-2}^2}e^{\alpha\phi_0} \coma\\
			\frac{1}{2}\delta(\delta-1)U_0\rho^{\delta-2}+\frac{D-2}{2}\delta\lambda U_0 \rho^{\delta-2}+\frac{2}{D-2}\Lambda_\AdSD{D}&=&\frac{D-3}{D-2}\frac{g_\UoD{D}^2\mc{Q}^2}{M_\PlD{D}^{D-2}\omega_{D-2}^2}\frac{e^{\alpha\phi_0}\rho^{\alpha A}}{R_0^{2(D-2)}\rho^{2(D-2)\lambda}}\fstop
		\end{array}\right.$} \eeq
	By matching the coefficients and powers, the parameters $a,A,\delta$ are related by the following equation
	\beq
	\frac{\alpha}{A}=\frac{(D-3)}{D-2}\frac{[\delta+(D-2)\lambda-1]}{[\frac{1}{2}\delta(\delta-1)+\frac{D-2}{2}\delta\lambda]}\fstop
	\eeq
	The final solution is found to be\footnote{One can check that the near-horizon solution obtained here is consistent with the flat space EMD black hole solution in \cite{Heidenreich:2015nta} upon setting $\Lambda=0$ and making the change of coordinates $r\rightarrow r_++\xi^p$, where $p=\frac{2}{\gamma_1+\gamma_2}=\frac{\gamma}{\frac{D-3}{D-2}+\frac{\alpha^2}{2(D-3)}}$.}
	\beq\label{eq:Alambdadeltacond}
	A=\frac{\alpha}{\frac{D-3}{D-2}+\frac{\alpha^2}{2(D-3)}}\coma \lambda = \frac{\alpha^2}{\frac{2(D-3)^2}{D-2}+\alpha^2}\coma \delta = \frac{2}{1+\frac{D-2}{(D-3)^2}\alpha^2}\fstop
	\eeq
	
	\subsubsection{Force Condition with Singular Extremal Solutions}
	\label{app:RFCforsingularEMDAdS}
	
	With the local solution \eqref{eq:local_expansion_singular} and \eqref{eq:Alambdadeltacond} at hand, we proceed with evaluating the radial force \eqref{eq:general_forceEMD}. Inserting the expansion, we obtain
	\beq \bal
	F& =-U_0\left(\frac{\delta}{2} +\frac{\alpha}{2} A\right) \rho^{\delta-1}+\left|\frac{\mf{q}}{\mf{m}}\right|e^{-\frac{\alpha}{2}\phi_0}\sqrt{U_0}\rho^{-\frac{\alpha}{2} A+\frac{\delta}{2}}\frac{g_\UoD{D}^2\mc{Q}}{\omega_{D-2}}\frac{\rho^{\alpha A} e^{\alpha\phi_0}}{R_0^{D-2}\rho^{\lambda(D-2)}}\\
	&=-U_0\left(\frac{\delta}{2}+\frac{\alpha}{2} A\right)\rho^{\delta-1}+\left|\frac{\mf{q}}{\mf{m}}\right|\frac{g_\UoD{D}^2\mc{Q}}{\omega_{D-2}}\frac{\sqrt{U_0}}{R_0^{D-2}}e^{\frac{\alpha}{2}\phi_0}\rho^{\frac{\delta}{2}+\frac{\alpha}{2}A-\lambda(D-2)}\fstop
	\eal \eeq
	Despite being divergent, the scaling of the forces with respect to $\rho$ are the same. Therefore, the repulsive condition can be obtained simply by comparing the coefficients,
	\beq
	\frac{\mf{q}}{\mf{m}}\ge\frac{U_0(\frac{\delta}{2}+\frac{\alpha}{2} A)}{\frac{g_\UoD{D}^2\mc{Q}}{\omega_{D-2}}\frac{\sqrt{U_0}}{R_0^{D-2}}e^{\frac{\alpha}{2}\phi_0}}\fstop
	\eeq
	The RHS is simplified using \eqref{eq:singular_equations}
	\beq \bal
	\text{RHS}&=\frac{g_\UoD{D}\mc{Q}}{M_\PlD{D}^{\frac{D-2}{2}}\omega_{D-2}}\sqrt{\frac{\alpha e^{\alpha\phi_0}}{A R^{2(D-2)}[1+\alpha A-(D-2)\lambda]}}\frac{(\frac{\delta}{2}+\frac{\alpha}{2} A)R_0^{D-2}}{\frac{g_\UoD{D}^2\mc{Q}}{\omega_{D-2}}+e^{\frac{\alpha}{2}+\frac{\alpha}{2}}}\\
	&= \sqrt{\frac{\alpha}{A}}\frac{1}{M_\PlD{D}^{\frac{D-2}{2}}g_\UoD{D}}\sqrt{1+\alpha A-(D-2)\lambda}\fstop
	\eal \eeq
	The condition on the charge-to-mass ratio then reads
	\beq \label{eq:pre-bound}
	\frac{M_\PlD{D}^{D-2}g_\UoD{D}^2\mf{q}^2}{\mf{m}^2}\ge  \frac{\alpha(1+\alpha A-(D-2)\lambda)}{A}\fstop
	\eeq
	Finally, replacing $A$ and $\lambda$ according to \eqref{eq:Alambdadeltacond}, we arrive at
	\beq
	\frac{M_\PlD{D}^{D-2}g_\UoD{D}^2\mf{q}^2}{\mf{m}^2}\ge \gamma_\ttiny{D}(\alpha)\coma
	\eeq
	which coincides with the WGC bound obtained in flat space. The repulsive condition between a singular dilaton BH in AdS being the same as that in flat space is a consequence of unstabilized dilaton, which makes the BH singular and zero size --- the scale of the BH is too small to experience effects of the cosmological constant.
	
	\subsection{Stabilized Dilaton and Smooth Extremal Solutions}
	\label{app:stabilizedEMDAdS}
	
	Extremal solutions with a smooth horizon can exist if the dilaton is stabilized near the black hole horizon, which can be achieved by introducing a dilaton mass term to the action \eqref{eq:EMDLambda_action}. This modification leads to an additional source term for the scalar equation in \eqref{eq:field_eqns}, i.e.
	\begin{equation}\label{eq:dilatonEOMmassive}
		\frac{1}{R^{D-2}}\frac{d}{dr}\left(R^{D-2}U\phi'\right)+\mu^2\phi 
		+2\beta \Lambda_\AdSD{D} e^{\beta\phi}  =  \frac{\alpha}{R^{2(D-2)}}  \frac{g_\UoD{D}^2\mc{Q}^2}{M_\PlD{D}^{D-2}\omega_{D-2}^2} e^{\alpha\phi}\coma
	\end{equation}
	while leaving the other equations untouched. The dilaton potential can be read-off as
	\begin{equation}
		\label{eq:dilaton_potential-appendix}
		V(\phi)=\frac{1}{2}\mu^2\phi^2-2(-\Lambda_\AdSD{D}) e^{\beta\phi}-\frac{1}{R^{2(d-2)}}\frac{g_\UoD{D}^2\mc{Q}^2}{M_\PlD{D}^{D-2}\omega_{D-2}^2} e^{\alpha\phi}\fstop
	\end{equation}
	This potential has a minimum when $\mu$ is sufficiently large, permitting regular solutions with the following local behavior of the fields near the extremal horizon
	\beq \label{eq:regular_solution_expansion}
	\renewcommand{\arraystretch}{1.7}\left.
	\begin{array}{rcl}
		\phi(r) &=&\phi_0+\phi_1\rho+\cdots\coma\\
		R(r)&=&R_0+R_1\rho+\cdots\coma\\
		U(r)&=&U_2\rho^2+\cdots\coma\\
	\end{array}\right. \eeq
	where, again, $\rho=\frac{r-r_h}{\ell_0}$. Inserting this ansatz into the field equations, we extract the following two equations 
	\beq \label{eq:local_smooth}\renewcommand{\arraystretch}{1.7}\left.
	\begin{array}{rcl}
		U_2+\frac{2}{D-2}\Lambda_\AdSD{D} e^{\beta\phi_0}&=&\frac{D-3}{D-2}\frac{g_\UoD{D}^2\mc{Q}^2}{M_\PlD{D}^{D-2}\omega_{D-2}^2} \frac{e^{\alpha\phi_0}}{R_0^{2(D-2)}}\coma\\
		U_2-\frac{(D-2)(D-3)}{2R_0^2}+\frac{D}{D-2}\Lambda_\AdSD{D} e^{\beta\phi_0}&=&\frac{D-4}{2(D-2)}\frac{g_\UoD{D}^2\mc{Q}^2}{M_\PlD{D}^{D-2}\omega_{D-2}^2}\frac{e^{\alpha\phi_0}}{R_0^{2(D-2)}}\coma\\
	\end{array}\right. \eeq
	which we then sum to relate the horizon dilaton value, the black hole charge and the black hole size
	\beq\label{eq:LambdaChargeSizeRel}
	-2\Lambda_\AdSD{D} e^{\beta\phi_0}+\frac{(D-2)(D-3)}{R_0^2}=\frac{g_\UoD{D}^2\mc{Q}^2}{M_\PlD{D}^{D-2}\omega_{D-2}^2}\frac{e^{\alpha\phi_0}}{R_0^{2(D-2)}}\fstop
	\eeq
	Meanwhile, the dilaton value at the horizon is fixed by 
	\beq
	\mu^2\phi_0-2\beta(-\Lambda_\AdSD{D})e^{\beta\phi_0}- \alpha\frac{g_\UoD{D}^2\mc{Q}^2}{M_\PlD{D}^{D-2}\omega_{D-2}^2}\frac{e^{\alpha\phi_0}}{R_0^{2(D-2)}}=0\fstop
	\eeq
	
	The smooth solutions are implicitly provided by the two transcendental equations above. As it turns out, the force condition can be expressed in a simple form with only the implicit local solution, which will be discussed in the following subsection.
	
	\subsubsection{Force Condition for Smooth Extremal Solutions}
	\label{app:RFCforstabilizedEMDAdS}
	
	To evaluate the force near the horizon of the regular solution, we again refer to \eqref{eq:general_forceEMD}. Applying \eqref{eq:regular_solution_expansion} to the radial force density, we get
	\beq \bal
	F
	&=U\left(-\frac{U'}{2U}+  \left|\frac{\mf{q}}{\mf{m}}\right| e^{-\frac{\alpha}{2}\phi}\frac{A_{t}'}{\sqrt{U}}-\frac{\alpha}{2} \phi'\right)\\
	&\sim \frac{U}{\rho}\left(-1+  \left|\frac{\mf{q}}{\mf{m}}\right| e^{-\frac{\alpha}{2}\phi_0}\frac{F_{01}}{\sqrt{U_2}}\right)
	\eal \eeq
	at the leading order. Repulsiveness then implies
	\beq\label{eq:partialresult}
	\begin{split}
		\frac{\mf{q}^2}{\mf{m}^2}&\ge \left.U_2\left(\frac{g_\UoD{D}^2 \mc{Q}}{\omega_{D-2}}\frac{e^{\frac{\alpha}{2}\phi_0}}{R_0^{D-2}}\right)^{-2}\right|_{\rho\rightarrow 0}\\
		&\geq \left.U_2\frac{R_0^{2D-4}e^{-\alpha\phi_0}}{g_\UoD{D}^4\mc{Q}^2\omega_{D-2}^{-2}}\right|_{\rho\rightarrow 0}\fstop
	\end{split} \eeq
	With a bit of manipulation, using \cref{eq:local_smooth,eq:LambdaChargeSizeRel} and recalling that $\Lambda_\AdSD{D}=-\frac{(D-1)(D-2)}{\ell^2_\AdSD{D}}$, we find
	\beq
	\left.\frac{g_\UoD{D}^2\mc{Q}^2}{M_\PlD{D}^{D-2}\omega_{D-2}^2}\frac{e^{\alpha\phi_0}}{R_0^{2(D-3)}}\right|_{\rho\rightarrow 0}=(D-2)(D-3)+(D-1)(D-2)\eta^2 e^{\beta\phi_0}\coma
	\eeq
	where we have introduced $\eta=\frac{r_h}{\ell_\AdSD{D}}$. The final repulsive force condition \eqref{eq:partialresult} becomes 
	\beq \label{eq:stabilized_WGC}
	\frac{M_\PlD{D}^{D-2}g_\UoD{D}^2\mf{q}^2}{\mf{m}^2}\ge \gamma_\ttiny{D}(0)\frac{1+\left(\frac{D-1}{(D-3)^2}+\frac{D-1}{D-3}e^{\beta\phi_0}\right)\eta^2}{1+\frac{D-1}{D-3}\eta^2}\fstop
	\eeq
	This expression reduces to that for an RN black hole by taking $\eta\rightarrow 0$ and also reduces to the AdS WGC in \cite{Lin:2025wfe} by taking $\phi_0\rightarrow 0$.
	
	\section{A Tower Weak Gravity Conjecture by Repulsive Force in Flat Space}
	\label{app:tower}
	
	For the case of dilatonic KK black holes with two $\U(1)$ charges given by the action \eqref{eq:EMDreducedaction}, we introduce a particle that also carries both charges and whose mass is given by \eqref{eq:scalar_mass}:
	\beq
	S_p=\int d\tau\left[\mf{m}(\phi,\lambda)\sqrt{-g_{\mu\nu}\dot{\xi}^\mu\dot{\xi}^\nu}+(\mf{q}A_\mu +\mf{q}_{\KK}B_\mu)\dot{\xi}^\mu\right]\fstop
	\eeq
	The force \eqref{eq:general_forceEMD} now receives additional contributions
	\beq \bal
	F=&g_{00}^{-1}\Gamma^1_{00}-\frac{1}{2}g^{11}\partial_\phi\ln{\mf{m}^2(\phi,\lambda)}\phi'-\frac{1}{2}g^{11}\partial_\lambda\ln{\mf{m}^2(\phi,\lambda)}\lambda'\\
	&+\frac{\mf{q}}{\mf{m}(\phi,\lambda)}\frac{g^{11}}{\sqrt{-g_{00}}}F_{01}+\frac{\mf{q}_\KK}{\mf{m}(\phi,\lambda)}\frac{g^{11}}{\sqrt{-g_{00}}}H_{01}\fstop
	\eal \eeq
	We extract the common scaling of the forces near the horizon and define the rescaled force as
	\beq
	\hat{F}(r)\equiv h^{-1}(r)F(r)\coma
	\eeq
	where $h(r)=g^{11}(\ln{f})'$. The force can be computed separately according to their source, which we collect as follows
	\beq \label{eq:force_components} \renewcommand{\arraystretch}{1.7}\left.
	\begin{array}{rcl}
		\hat{F}_g&=&-\frac{1}{\gamma_\ttiny{D}(\alpha)}\frac{d-3}{d-2}\left(1+\hat{f}_\lambda^{-1}f^\frac{2}{\gamma_\ttiny{D}(\alpha)}\right)\coma\\
		\hat{F}_\phi&=&-\frac{\alpha}{2\gamma_\ttiny{D}(\alpha)}\partial_ \phi \ln{\mf{m}^2(\phi,\lambda)}\coma\\
		\hat{F}_\lambda&=&-\frac{1}{\gamma_\ttiny{D}(\alpha)}\left(\frac{1}{d-1}+f^\frac{2}{\gamma}\hat{f}_\lambda^{-1}\right)\partial_\lambda \ln{\mf{m}^2(\phi,\lambda)}\coma\\
		\hat{F}_{\U(1)}&=&\frac{g_\UoD{d}e^{\frac{\alpha}{2}\phi+\frac{\lambda}{2(d-2)}}\mf{q}M_\PlD{d}^\frac{d-2}{2}}{\sqrt{\gamma_\ttiny{D}(\alpha)}\mf{m}(\phi,\lambda)}\frac{u}{v}\hat{f}_\lambda^{-\frac{1}{2}}\coma\\
		\hat{F}_\KK&=&\frac{e_\KK e^{\frac{d-1}{2(d-2)}\lambda}\mf{q}_\KK M_\PlD{d}^\frac{d-2}{2}}{\sqrt{\gamma_\KK}\mf{m}(\phi,\lambda)}\frac{2}{\gamma_\ttiny{D}(\alpha)}\frac{u}{v}\hat{f}_\lambda^{-1}f^\frac{1}{\gamma_\ttiny{D}(\alpha)}\fstop
	\end{array}\right. \eeq
	We find the scalar forces by making use of the expression for the particle mass,
	\beq \renewcommand{\arraystretch}{1.7}\left.
	\begin{array}{rcl}
		\hat{F}_\phi&=&-\frac{\alpha^2}{2\gamma_\ttiny{D}(\alpha)}\frac{1}{1+\zeta^2}\coma\\
		\hat{F}_\lambda&=&-\frac{1}{\gamma_\ttiny{D}(\alpha)}\left(\frac{1}{d-1}+f^\frac{2}{\gamma_\ttiny{D}(\alpha)}\hat{f}_\lambda^{-1}\right)\left(\frac{1}{d-2}+\frac{\zeta^2}{1+\zeta^2}\right)\coma
	\end{array}\right. \eeq
	where $\zeta^2=e^{\lambda-\alpha\phi}\frac{\mf{q}_{\KK}}{\mf{m}^2r_{S^1}^2}$
	Some force contributions are parametrically small near the horizon as $f(r\rightarrow r_h)=0$ and $ \zeta\rightarrow \infty$. We can neglect these contributions obtaining a net force of
	\beq
	\left.\hat{F}\right|_{r\rightarrow r_h}=\left.\hat{F}_{\U(1)}\right|_{r\rightarrow r_h}-\frac{1}{\gamma_\ttiny{D}(\alpha)}\fstop
	\eeq
	In order for the total force to be repulsive close to the horizon, i.e. $\left.\hat{F}\right|_{r\rightarrow r_h}\geq 0$, we need to ensure that the Coulomb force is greater than $\gamma_\ttiny{D}(\alpha)^{-1}$. Manipulating $\left.\hat{F}_{\U(1)}\right|_{r\rightarrow r_h}$, we obtain that 
	\beq\label{eq:zDforKK}
	\frac{g_\UoD{d}e^{\frac{\alpha}{2}\phi+\frac{\lambda}{2(d-2)}}\mf{q}M_\PlD{d}^\frac{d-2}{2}}{\sqrt{\gamma_\ttiny{D}(\alpha)}\mf{m}(\phi,\lambda)}=\frac{\mf{z}_\ttiny{D}}{\sqrt{1+e^{\lambda-\alpha\phi}\frac{\mf{q}_\KK^2}{\mf{m}^2r_{S^1}^2}}}\geq \frac{1}{\gamma_\ttiny{D}(\alpha)}\coma
	\eeq
	or equivalently 
	\beq\label{eq:radiuscondition-appendix}
	\mf{m}^2 r_{S^1}^2\ge \frac{\frac{\mf{q}_{\KK}^2}{\gamma_\ttiny{D}(\alpha)^2}e^{\lambda-\alpha\phi}}{\mf{z}_\ttiny{D}^2-\frac{1}{\gamma_\ttiny{D}(\alpha)^2}}\fstop
	\eeq
	
	The presence of $\gamma_\ttiny{D}(\alpha)^{-2}$ on the RHS of \eqref{eq:radiuscondition-appendix} is due to the evaluation of the force at the horizon for a charged KK mode of the particle $\mathfrak{z}_\ttiny{D}$. The final requirement on the spectrum of the particle in $D$-dimension is determined by considering extremal black holes and particles without KK charge, i.e. $\mf{q}_\KK = 0$. This removes the divergence in $\zeta$ near the horizon, and the force contributions become
	\beq \renewcommand{\arraystretch}{1.7}\left.
	\begin{array}{rcl}
		\hat{F}_g&=&-\frac{1}{\gamma_\ttiny{D}(\alpha)}\frac{d-3}{d-2}\coma\\
		\hat{F}_\phi&=&-\frac{\alpha^2}{2\gamma_\ttiny{D}(\alpha)}\coma\\
		\hat{F}_\lambda&=&-\frac{1}{\gamma_\ttiny{D}(\alpha)}\frac{1}{d-1}\frac{1}{d-2}\coma
	\end{array}\right. \eeq
	so that the total force near the horizon is 
	\beq \bal
	\hat{F}&=\hat{F}_{U(1)}-\frac{1}{\gamma_\ttiny{D}(\alpha)}\left[\frac{d-3}{d-2}+\frac{\alpha^2}{2}+\frac{1}{(d-1)(d-2)}\right]\\
	&=\hat{F}_{U(1)}-1\coma
	\eal \eeq
	for any value of $\alpha$. This is consistent with Section \ref{sec:EMDinflatspace} where we showed that the requirement for the net force to be repulsive close to the horizon is to have a particle with $\mathfrak{z}_\ttiny{D}\geq 1$.

	\bibliographystyle{JHEP}
	\bibliography{references}
	
\end{document}